\documentclass[aps,prb,twocolumn,superscriptaddress]{revtex4-2}

\usepackage[colorlinks=true, allcolors=blue]{hyperref}
\usepackage{amsmath,amssymb,amsfonts,upgreek}
\usepackage{color}
\usepackage{graphicx}
\usepackage{setspace}
\usepackage{bm}
\usepackage[export]{adjustbox}
\usepackage{dsfont}
\usepackage{braket}
\usepackage{multirow}
\usepackage{changes}

\usepackage{soul}
\setulcolor{red}
\usepackage{comment}

\usepackage{tikz-feynman}

\definecolor{myred}{rgb}{0.8,0.1,0.1}
\definecolor{myblue}{rgb}{0.0,0.1,0.6}

\newcommand{\crea}[2]{ #1^{\dagger}_{\mathrm{\small{#2}}} }
\newcommand{\ann}[2]{ {#1}^{\mathstrut}_{\mathrm{\small{#2}}}}

\begin{document}

\title{Revised Tolmachev-Morel-Anderson pseudopotential \\
for layered conventional superconductors with nonlocal Coulomb interaction}

\author{M. Simonato}
\author{M. I. Katsnelson}
\author{M. R\"osner}
\affiliation{Institute for Molecules and Materials, Radboud University, 6525 AJ Nijmegen, the Netherlands}

\date{\today}

\begin{abstract}

We study the effects of static nonlocal Coulomb interactions in layered conventional superconductors and show that they generically suppress superconductivity and reduce the critical temperature. Although the nonlocal Coulomb interaction leads to a significant structure in the superconducting gap function, we find that most properties can be effectively described by means of an appropriately revised local Coulomb pseudopotential $\tilde{\mu}_C^*$, which is \emph{larger} than the commonly adopted retarded Tolmachev-Morel-Anderson pseudopotential $\mu^*_C$.
To understand this, we analyze the Bethe-Salpeter equation describing the screening of Coulomb interaction in the superconducting state and obtain an expression for $\tilde{\mu}_C^*$, which is valid in the presence of nonlocal Coulomb interactions. This analysis also reveals how the structure of the nonlocal Coulomb interaction weakens the screening effects from high-energy pair fluctuations and therefore yields larger values of the pseudopotential. Our findings are especially important for layered conventional superconductors with small Fermi energies and can be easily taken into account ab initio studies.
\end{abstract}

\maketitle

\section{Introduction}

Conventional superconductivity in the weak and strong electron-phonon coupling limit is well described within Migdal-Eliashberg (ME) theory \cite{osti_7354388,carbotte_properties_1990}, which accurately approximates superconducting properties around the Fermi level \cite{Allen_Mitrovich,vonsovsky_superconductivity_2011}. When both electron-phonon and electron-electron (Coulomb) interactions are present the corresponding expressions become, however, rather involving \cite{secchi_phonon-mediated_2020,anokhin_phonon-induced_1996}. Thus, even for the description of conventional superconductivity approximations are required that take the Coulomb repulsion in the pairing channel adequately into account. A conventional scheme to do so was independently introduced by Tolmachev \cite{Tol_1961} and by Morel and Anderson \cite{morel_calculation_1962} yielding the famous Tolmachev-Morel-Anderson (TMA) \emph{local} retarded Coulomb pseudopotential $\mu^*_C$. Using appropriate energy scales, the TMA expression provides good estimates for $\mu^*_C$ which, together with phonon properties from first principles, yield good agreement with experimental data for many elemental bulk superconductors and their alloys \cite{bauer_retardation_2013,carbotte_properties_1990}.

Beside the successful descriptions obtained with the conventional TMA approach, there are situations in which the details of the Coulomb interaction and its screening become important and need to be carefully considered. This includes the dynamics of the screening, which can lead to plasmonic contributions to superconductivity \cite{wang_origin_2022,bill_electronic_2003,rietschel_role_1983,richardson_effective_1997,takada_s-_1993,akashi_development_2013}, and the possible nonlocal character of the electron-electron interaction. The latter can be crucially important in case of discorded systems \cite{anderson_theory_1983,bulaevskii_anderson_1985} or when the screening radius is larger than the correlation length, i.e. the Cooper pair radius. This defines the difference between superconducting bulk metals, where the screening radius is of the order of the lattice constant such that the Coulomb interaction can be reasonably well described by a constant local Hubbard $U$, and slightly-doped semiconductors, where the screening radius can be very large, which can become even more problematic in 2D as the screening properties of 3D and 2D electron gases are essentially different \cite{ando_electronic_1982,kotov_electron-electron_2012, katsnelson_physics_2020}.

For layered superconductors with reduced screening and hence naturally enhanced nonlocal Coulomb interactions it is thus a priori not clear whether the conventional local TMA pseudopotential is still a valid description. Nevertheless, it has been regularly applied to study superconductors in 2D \cite{rosner_phase_2014, schonhoff_interplay_2016,  margine_electron-phonon_2016, bekaert_evolution_2017, jishi_theoretical_nodate, lian_unveiling_2018, zheng_electron-phonon_2019,  bekaert_hydrogen-induced_2019,  lian_intrinsic_2022, singh_high-temperature_2022, sevik_high-temperature_2022}. This together with numerous recent experiments on layered superconductors \cite{he_phase_2013,liu_electronic_2012, wang_interface-induced_2012, reyren_superconducting_2007, guo_superconductivity_2004}, motivates us to study here the effects of long-range Coulomb interactions in conventional 2D superconductivity (dynamical Coulomb effects are discussed elsewhere \cite{wang_origin_2022, yann}). We therefore use the static Thomas-Fermi approximation of the Coulomb interaction, which coincides with the more accurate random phase approximation for $q < 2k_F$ for both parabolic \cite{ando_electronic_1982} and Dirac \cite{kotov_electron-electron_2012, katsnelson_physics_2020} spectra.

We find that a local approximation of the Coulomb interaction is actually still suitable to describe the relevant superconducting quantities even in the presence of long-range nonlocal static Coulomb interactions. We show, however, that the widely adopted TMA $\mu_C^*$ strongly overestimates the superconducting gap function at the Fermi level and thus the critical transition temperature. In the case of nonlocal Coulomb interactions, the screening effects resulting from virtual pair fluctuations at high energies \emph{above} the Fermi level are strongly suppressed. This leads to a larger value of the screened TMA pseudopotential $\tilde{\mu}_C^*$, which in turn yields smaller superconducting gaps and reduced critical temperatures $T_c$. We derive a generalised expression for the evaluation of the new $\tilde{\mu}_C^*$, which takes into account the nonlocality of the static Coulomb interaction. Our expression elucidates quantitatively how the inverse screening length $\Lambda$ and the chemical potential control the parameter $\tilde{\mu}_C^*$, and thus the critical temperature $T_c$. From this analysis it also emerges that the nonlocality of the Coulomb interaction is crucial for layered SCs with low carrier-density (small Fermi energies) and small effective masses.

The paper is organized as follows. In section II we introduce the extended BCS model in 2D to account for the nonlocal Coulomb interaction. In section III we present our main results obtained from numerical solutions of the extended gap equation. In section IV we analyse the Bethe-Salpeter equation describing the screening of the Coulomb interaction within the superconducting state due to virtual pair fluctuations. We derive the generalised expression for $\tilde{\mu}_C^*$ which we compare to the numerical data. Section V summarises our findings and highlights in which regime they are most important. In appendix \ref{app:sol} we furthermore show that the expression for $\tilde{\mu}_C^*$ can also be derived from an analytical approximate solution of the gap equation in the presence of nonlocal Coulomb interactions.

\section{Model Definitions and Properties}

To study the effects of static nonlocal Coulomb interactions to the superconducting properties of a layered system, we solve the following Hamiltonian
\begin{equation}
    H = \sum_{k,\sigma}  \xi_k \crea c{k\sigma} \ann c{k\sigma}
    + \sum_{k,k'}[ - g_{kk'}  + V_{kk'} ] 
    \crea c{k\uparrow} \crea c{-k\downarrow}  
    \ann c{-k'\downarrow} \ann c{k' \uparrow}
\end{equation}
within mean-field BCS theory, where $\ann c{k\sigma} $ $(\crea c{k\sigma})$ denotes the annihilation (creation) of a an electron with spin $\sigma$ and momentum $k$. $\xi_k = \frac{\hbar^2 k^2}{2m^*} - \mu$ is a 2D electron gas dispersion with the chemical potential $\mu$ and the effective mass $m^*$. $g_{kk'}$ and $V_{kk'}$ describe the effective attractive interaction mediated by phonons and the static Coulomb repulsion between electrons, respectively. For the phonon mediated attraction we use the BCS model \cite{bardeen_theory_1957}
\begin{equation}
    g_{kk'} =
    \begin{cases}  
        g \quad  \text{for } |\xi_k|  <\omega_D \text{ and } |\xi_{k'}|  <\omega_D  \\
        0  \quad \text{elsewhere}
    \end{cases},
\end{equation}
which allows for electron paring within the Debye energy $\omega_D$ around the Fermi level. To understand the effects of nonlocal Coulomb repulsion we employ the static Thomas-Fermi (TF) approximation for the Coulomb interaction kernel $V_{kk'}$  
\begin{equation}
    V_{kk'} = \frac{2\pi e^2}{\Omega} \frac{1}{\varepsilon |k-k'| + \Lambda},
\end{equation}
where $\Omega$ denotes the normalization area, $e$ the electron charge, and $\varepsilon$ a homogeneous local screening. We allow the TF wavevector $\Lambda$ to be a free parameter as it serves as a measure for the nonlocality. After mean-field decoupling we obtain the gap equation
\begin{equation}
    \Delta_k = - \sum_{k'} [-g_{kk'} + V_{kk'} ]  
                \frac{\tanh{(\beta E_{k'}/2)}}{2 E_{k'}}\Delta_{k'},
    \label{eq:gap_eq}
\end{equation}
where $E_k =\sqrt{\xi_k^2 + \Delta_k^2}$ is the Bogoliubov dispersion relation and $\beta$ the inverse temperature. In the presence of the nonlocal Coulomb interaction there are no analytical solutions known and we need to solve the gap function numerically. The common approximation to gain anyway analytical insights into the problem is by projecting the Coulomb kernel onto the Fermi surface via a double average of the form \cite{morel_calculation_1962, Tol_1961}
\begin{equation}
    \mu_C = \braket{\braket{ V_{kk'} }}_{\textrm{FS}} 
          = \frac{1}{\rho_0} \sum_{k,k'} V_{kk'} 
            \delta{(\xi_k) } \delta{(\xi_{k'})}
          =: \rho_0 U, 
    \label{eq:muC_2FS}
\end{equation}
where $\rho_0$ denotes the normal density of states at the Fermi level and $U$ describes an effective local interaction. We denote the gap function obtained with this local approximation by $\Delta_k^L$, which can be analytically derived yielding \cite{morel_calculation_1962, coleman_introduction_2015}
\begin{equation}
    \Delta_k^L = 
    \begin{cases}  
        \Delta_1^L \quad \text{for } |\xi_k| <\omega_D \text{ and } |\xi_{k'}| < \omega_D  \\
       \Delta_2^L \quad \text{elsewhere}
    \end{cases},
    \label{eq:deltaL}
\end{equation}
with
\begin{equation} 
    \Delta_1^L = \frac{\omega_D}{\sinh( \frac{1}{\lambda-\mu^*_C})} \quad \text{and} \quad \Delta_2^L = - \frac{\mu^*_C}{\lambda - \mu^*_C} \Delta_1^L \, .\label{eq:deltaLSol}
\end{equation}
Here $\lambda = \rho_0 g$ is the effective electron-phonon coupling strength and $\mu^*_C$ is the retarded Tolmachev-Morel-Anderson (TMA) pseudopotential defined by
\begin{equation}
    \mu^*_C = \frac{\mu_C}{1 + \mu_C \log\frac{D}{\omega_D}},
    \label{eq:muStar}
\end{equation} 
where $D$ is the electron cut-off energy, which is typically of the order of the electron bandwidth \cite{morel_calculation_1962, coleman_introduction_2015}. The retarded pseudopotential $\mu^*_C$ is always smaller than the bare $\mu_C$ due to screening effects from virtual pair fluctuations at energies between $\omega_D$ and $D$. Thus, even for $\lambda - \mu_C < 0 $ we can have $\lambda - \mu^*_C > 0$, i.e. a superconducting solution can exist. This underlines the importance of screening of the Coulomb repulsion by the high energy degrees of freedom in this problem.

In the following we explore the validity of the local approximation of the Coulomb repulsion in the presence of a nonlocal Coulomb kernel and clarify the role of the nonlocality to the superconducting gap function, critical temperature $T_c$, and spectral function.

\section{Gap Function Structure with Local and Nonlocal Coulomb Kernels}

\begin{figure}[h]
    \centering
    \includegraphics[width =1.0\linewidth, keepaspectratio]{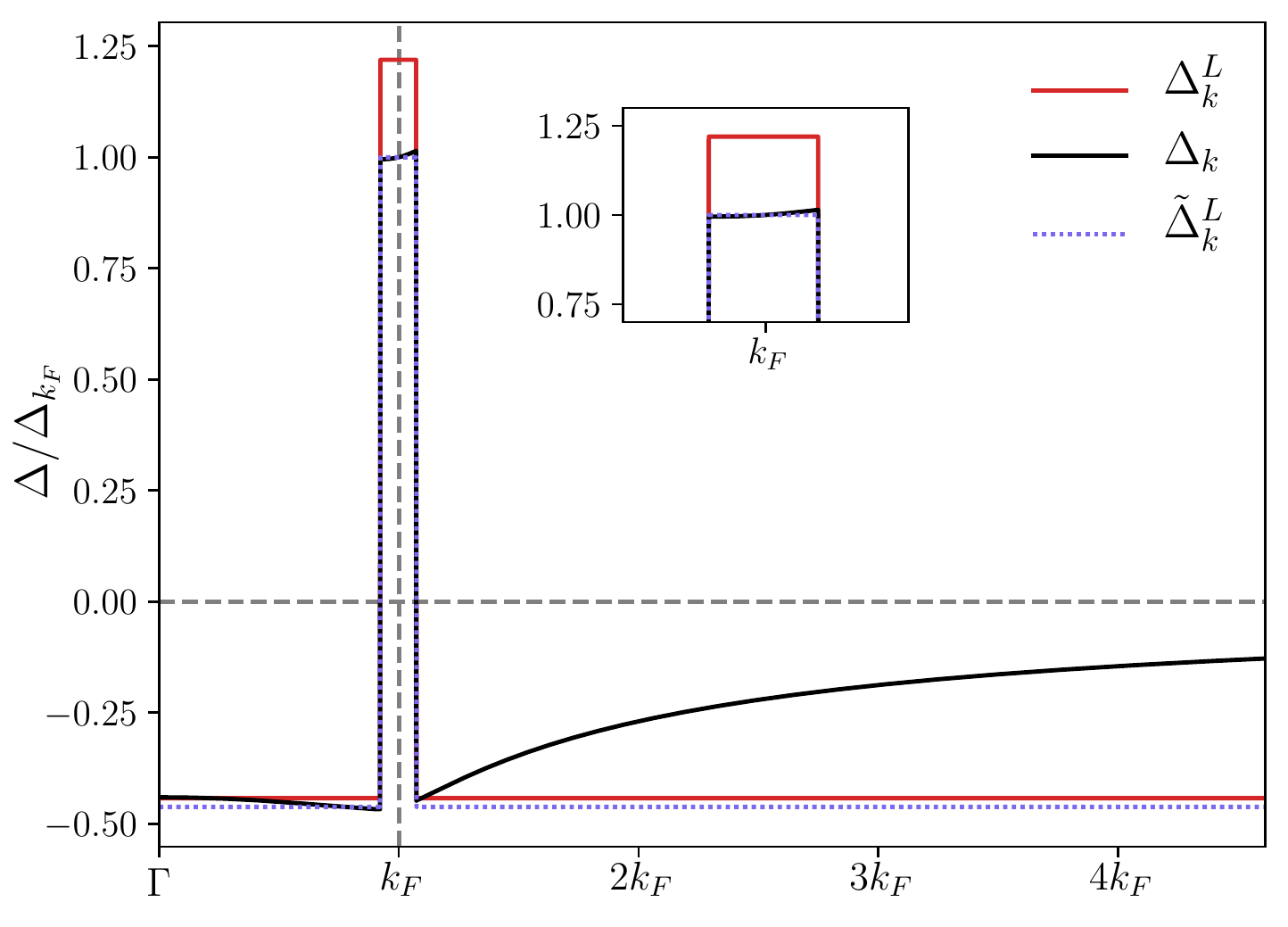}
    \caption{Gap functions $\Delta_k$ and $\Delta^L_k$ obtained by solving the gap equation at $T=0$ using the full non-local Coulomb kernel and a local one utilizing the conventional TMA pseudopotential $\mu_C^*$, respectively. $\tilde{\Delta}^L_k$ represents the gap function with a local Coulomb kernel fitted to reproduce the $\Delta_k$ at $k_F$.}
    \label{fig:gap1}
\end{figure}

In Fig.~\ref{fig:gap1} we show the gap function $\Delta_k$ as obtained from numerically solving Eq.~(\ref{eq:gap_eq}) in polar coordinates (see appendix \ref{app:polar}) with the full nonlocal Coulomb kernel $V_{kk'}$ and using $m^*/m_e=0.5$, $\mu= 0.5$ eV, $ \lambda = 0.5 $, $\omega_D=75$ meV, $\Lambda= 1.5 $ \AA$^{-1}$, and $\varepsilon= 10$. The cut-off in $k$-space is set to the Debye wavevector $k_D = \sqrt{4 \pi}/a$ with $a = 3$ \AA. For comparison we also show the analytical results for the local approximation $\Delta_k^L$ from Eq.~(\ref{eq:deltaL}) using the Fermi-surface-averaged local Coulomb interaction from Eq.~(\ref{eq:muC_2FS}) yielding $\mu^*_C= 0.135 $ via Eq.~(\ref{eq:muStar}). From this comparison we find two important differences: (1) Similar to $\Delta_k^L$, $\Delta_k$ also exhibits a separation into positive (low-energy) and negative (high-energy) components, but the nonlocality of the Coulomb repulsion induces a significant structure in the negative part (especially towards large $k$ away from $k_F$) as well as a bending in the positive parts around $k_F$. (2) The conventional local approximation significantly overestimates the gap functions at the Fermi level, i.e., $\Delta_{k_F}^L > \Delta_{k_F}$, and also underestimates the negative parts.

\begin{figure}[h]
    \centering
    \includegraphics[width =\linewidth, keepaspectratio]{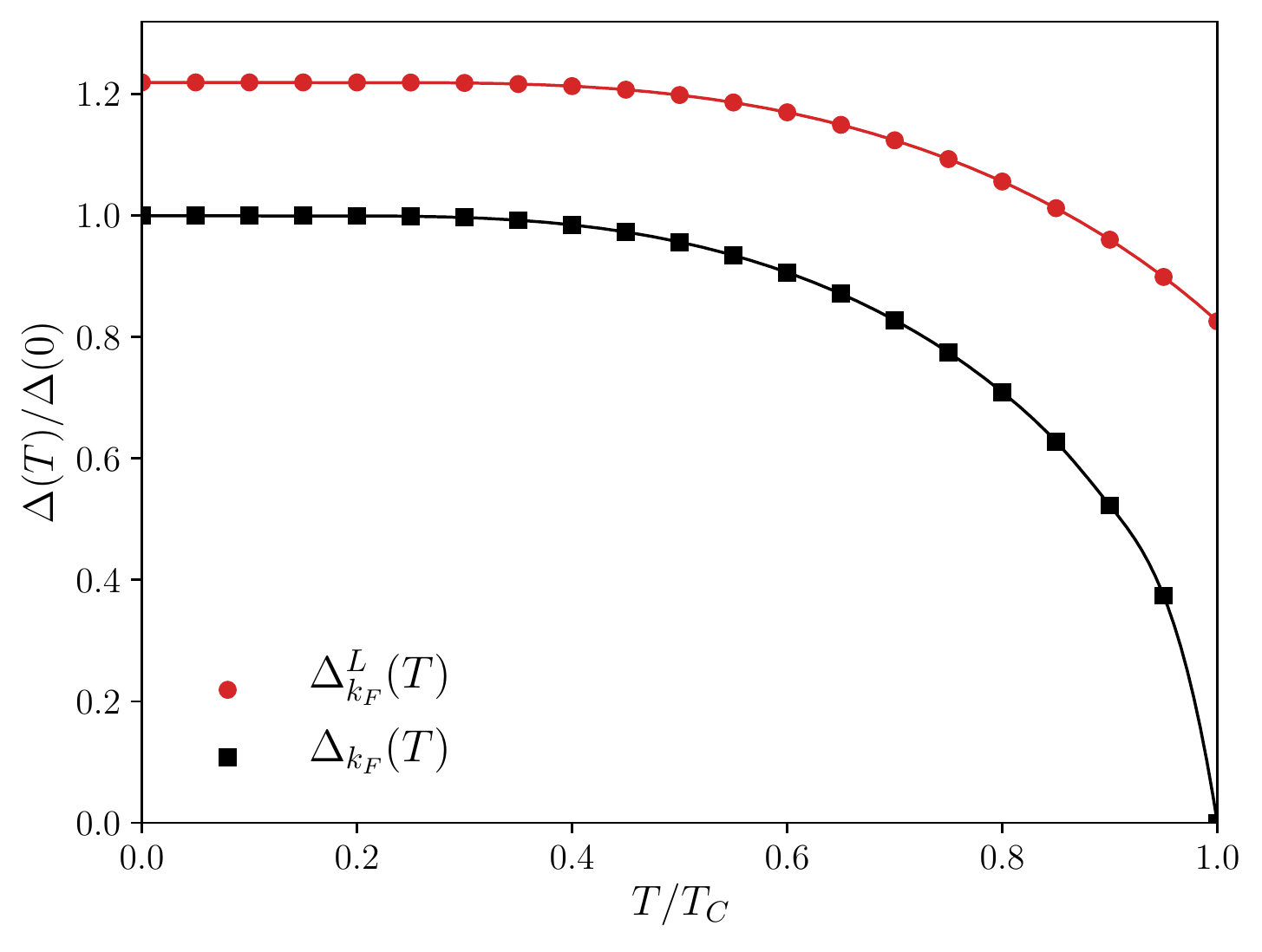}
    \caption{Gap functions at $k_F$ as function of temperature $T$. The temperature axis is normalized to the critical temperature $T_c$ as predicted with BCS relation using $\Delta_{k_F}(0)$. }
    \label{fig:gapT}
\end{figure}

In Fig.~\ref{fig:gapT} we show both gap functions evaluated at the Fermi level $\Delta^L_{k_F}(T)$ and $\Delta_{k_F}(T)$ as functions of the temperature $T$. The data shows that the the conventional local approximation also significantly overestimates the true critical temperature $T_c$ of the nonlocal problem. Furthermore and even more important, we find that in the full nonlocal case $T_c$ is related to zero-temperature value of the gap at the Fermi level $\Delta_{k_F}(0)$ by the BCS ratio $T_c = \frac{2 \Delta_{k_F}(0)}{3.35}$.

\begin{figure}[htb]
    \centering
    \includegraphics[width =\linewidth, keepaspectratio]{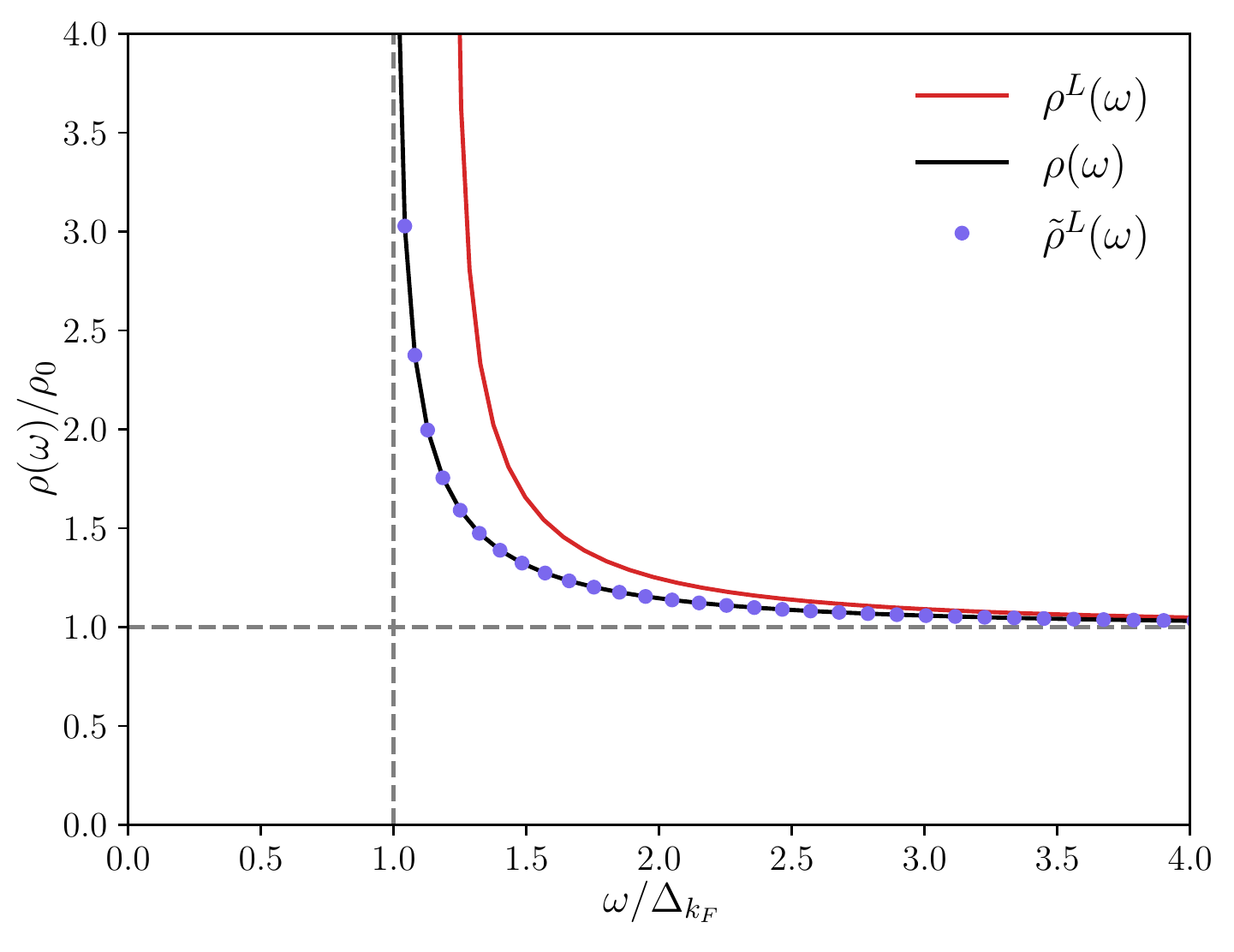}
    \caption{Comparison of SC spectral functions $\rho(\omega)$, $\rho^L(\omega)$, and $\tilde{\rho}^L(\omega)$ resulting from $\Delta_{k}$, $\Delta^L_{k}$, and $\tilde{\Delta}^L_{k}$, respectively.}
    \label{fig:spectral}
\end{figure}

This observation motivates us to \emph{fit} the gap function $\Delta_{k}$ of the nonlocal interaction model with a gap function $\tilde{\Delta}^{L}$ of the form of Eq.~(\ref{eq:deltaL}), i.e. using a local Coulomb interaction.
This gap $\tilde{\Delta}^{L}$ is constructed by adjusting $\mu^*_C$ in Eq.~(\ref{eq:deltaL}) such that $ \tilde{\Delta}^L_{k_F} =\Delta_{k_F} $ holds. This yields a significantly \emph{enhanced} $\tilde{\mu}_C^* = 0.157$ as compared to $\mu^*_C= 0.135 $  calculated from Eq.~(\ref{eq:muStar}). The resulting effective local model gap function at $T=0$ is shown in Fig.~(\ref{fig:gap1}), which yields by definition the same $T_c$ as the full model. To further investigate the quality of this effective local model, which disregards all curvature of $\Delta_{k}$, we calculate the interacting spectral functions $\rho(\omega) = \int dk \delta(\omega - E_k)$ for all three gap functions ($\Delta_{k}$, $\Delta^L_{k}$, and $\tilde{\Delta}^L_{k}$) and show them in Fig.~\ref{fig:spectral}.

We find that the full spectral function $\rho(\omega)$ can be accurately approximated by the spectral function $\tilde{\rho}^L(\omega)$ obtained from the fitted local interaction model. Thus, the curvature of $\Delta_{k}$ does not have a major impact to the spectral function, which has a two-fold reason: the bending in $\Delta_k$ within the low-energy region around $k_F$ is rather small and does not affect strongly the coherence peaks. Secondly, as soon as $|\xi| > \omega_D$ the detailed structure of $\Delta_k \ll \omega_D$ does not affect the Bogoliubov dispersion $E= \sqrt{\xi^2 + \Delta^2}$ anymore. The negative component of $\Delta_k$ thus leaves no significant trace in the spectral function.

This analysis shows that within the BCS framework the value of the gap function at $k_F$ is sufficient for the evaluation of the relevant SC quantities in layered materials. 
However, the data also shows that in this generic model the commonly adopted TMA approach always overestimates the value of the gap function at the Fermi level and therefore overestimates the critical temperature $T_c$ as it underestimates $\mu^*_C$. The nonlocal Coulomb interactions thus reduces the gap function at the Fermi level. 

\begin{figure}[htb]
    \centering
    \includegraphics[width =\linewidth, keepaspectratio]{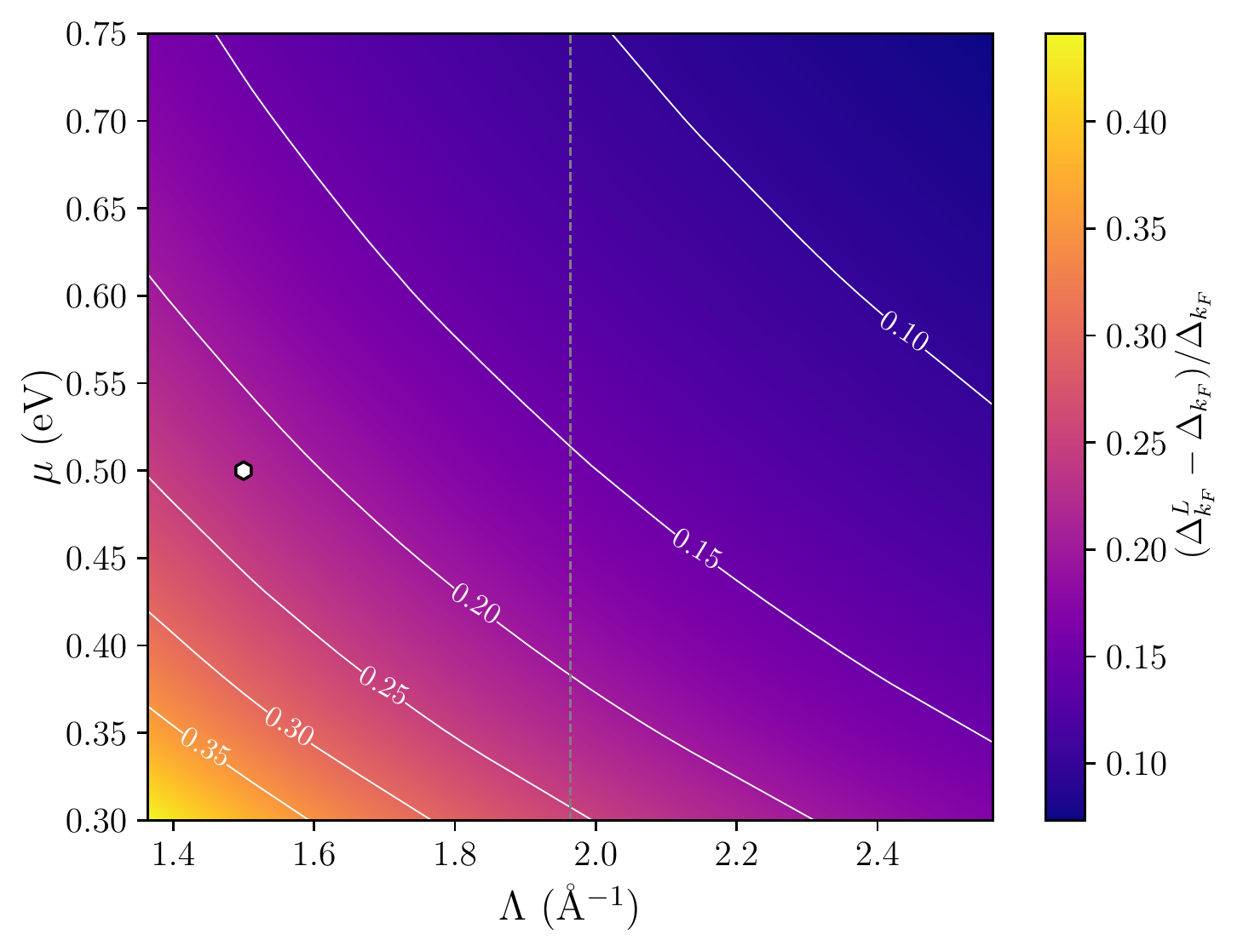}
    \caption{Relative difference between $\Delta_{k_F}^L$ and $\Delta_{k_F}$ as a function of inverse screening length $\Lambda$ and chemical potential $\mu$. The vertical dashed line denotes the Thomas-Fermi inverse screening length  $\Lambda_{TF} = 4\pi e^2 \rho_0 / \Omega$ and the white marker denotes the parameters $(\Lambda, \mu)$ as used for the data shown in Figs.~\ref{fig:gap1},\ref{fig:gapT},\ref{fig:spectral}}
    \label{fig:rel_gap}
\end{figure}

This is indeed generic as depicted in Fig.~\ref{fig:rel_gap} where we show the relative difference between the full $\Delta_{k_F}$ and the conventional local TMA approximation $\Delta^L_{k_F}$. This deviation is controlled by the Thomas-Fermi inverse length $\Lambda$ as well as by the chemical potential $\mu$. As $\Lambda$ decreases, the nonlocality in the Coulomb interaction becomes stronger and thus we find larger differences between $\Delta_{k_F}$ and $\Delta^L_{k_F}$. Interestingly, the deviation is also found to increase for smaller chemical potentials. In the next section, the analysis of the Bethe-Salpeter equation will reveal the physical reasons determining this behaviour and provides an accurate quantitative description.

\section{Bethe-Salpeter equation and revised Tolmachev-Morel-Anderson potential}

From the previous analysis we understand that a local Coulomb interaction model can accurately reproduce the gap function of the full nonlocal Coulomb model at $k_F$, which controls most relevant SC quantities. The possibility to employ a local Coulomb kernel is a significant simplification of the problem, as the gap function at the Fermi level admits in this case a simple analytical solution Eq.~(\ref{eq:deltaL}). We therefore aim to understand the nature of the parameter $\tilde{\mu}_C^*$, which is responsible for the deviation between $\Delta_{k_F}$ and $\tilde{\Delta}^L_{k_F}$.

The discrepancy between $\mu^*_C$ and the correct value $\tilde{\mu}_C^*$ is to be attributed to the role of nonlocality of the Coulomb interaction in the retardation effects. To understand this, we analyse the Bethe-Salpeter equation (BSE) for the case of a nonlocal interactions:
\begin{equation*}
\feynmandiagram [ horizontal =i1 to a, horizontal= c to d, small, baseline=(current bounding box.center)] { 
 i1 -- [fermion, edge label= \( k \)] a  -- [fermion, edge label= \( k' \)] f1,
 i2 -- [fermion, edge label'= \( -k \)] c -- [fermion, edge label'= \( -k' \)] f2,
 a -- [ photon, very thick, edge label' =\(  \textbf{W}_{kk'} \), ]c,
}; 
= 
\feynmandiagram [small, baseline=(current bounding box.center)] { 
 i1 -- [fermion, edge label= \( k \),] a -- [fermion, edge label= \( k' \)] f1,
 i2 -- [fermion, edge label'= \( -k \)] c -- [fermion, edge label'= \( -k' \)] f2,
 a -- [photon, edge label' =\( V_{kk'} \), thin ] c,
};\\
-
\feynmandiagram [small, baseline=(current bounding box.center)] { 
 i1 -- [fermion, edge label= \( k \)] a -- [fermion, edge label= \( q \)] t1 -- [fermion, edge label= \( k' \)] f1,
 i2 -- [fermion, edge label'= \( -k \)] c -- [fermion, edge label'= \( -q \)] r1 --[fermion, edge label'= \( -k' \)] f2,
 a -- [ photon, edge label' =\(  V_{kq} \) ] c,
 t1--[ photon, edge label = \(  \textbf{W}_{qk'} \), very thick]r1 ,
}; 
\end{equation*}

After performing the Matsubara summation and evaluating at $T=0$ the BSE equation reads
\begin{equation}
    W_{kk'} = V_{kk'} - \sum_{q \in \bar{\chi}} V_{kq}\frac{1}{2\xi_q} W_{qk'},
\end{equation}
where $W_{kk'}$ denotes the Coulomb potential screened by the virtual pair fluctuations, and the summation only runs in the high energy region $\bar{\chi}$, where $|\xi_q|>\omega_D$. For a local potential $V_{kk'}=U$ one obtains the conventional retarded TMA potential $\mu^*_C$ in the form of Eq.~(\ref{eq:muStar}).

From our previous analysis we understand that a single parameter $\tilde{\mu}_C^*$ is capable to encode the relevant retardation effects affecting the energy gap \emph{at the Fermi level}. To correctly account for this it is therefore necessary to evaluate the retarded (nonlocal) potential $W_{kk'}$ first, and only subsequently project it onto the Fermi surface. This procedure is depicted schematically in Fig.~\ref{fig:BSEApproach}. By commuting the order of operations, which is conventionally employed to estimate $\mu^*_C$, we ensure that the nonlocal screening effects due to high energy virtual pair fluctuations are evaluated \emph{before} the potential is projected onto the Fermi surface.
\begin{figure}[h]
    \textbf{Conventional TMA approach:}
    \begin{equation*}
        V_{kk'} \rightarrow \fcolorbox{blue}{white}{\text{FS  projection}} \rightarrow \mu_C \rightarrow 
        \fcolorbox{red}{white}{\text{BSE}} \rightarrow \mu^*_C
    \end{equation*}
    \textbf{Revised approach:}
    \begin{equation*}
        V_{kk'} \rightarrow  
         \fcolorbox{red}{white}{\text{BSE}}
        \rightarrow W_{kk'}  
         \rightarrow \fcolorbox{blue}{white}{\text{FS  projection}}
        \rightarrow \tilde{\mu}_C^*
    \end{equation*}
    \caption{Schematic representation of the conventional TMA approach in contrast to our revised approach to evaluate the retarded potential.}
    \label{fig:BSEApproach}
\end{figure}

\begin{figure}[htb]
    \centering
    \includegraphics[width =\linewidth, keepaspectratio]{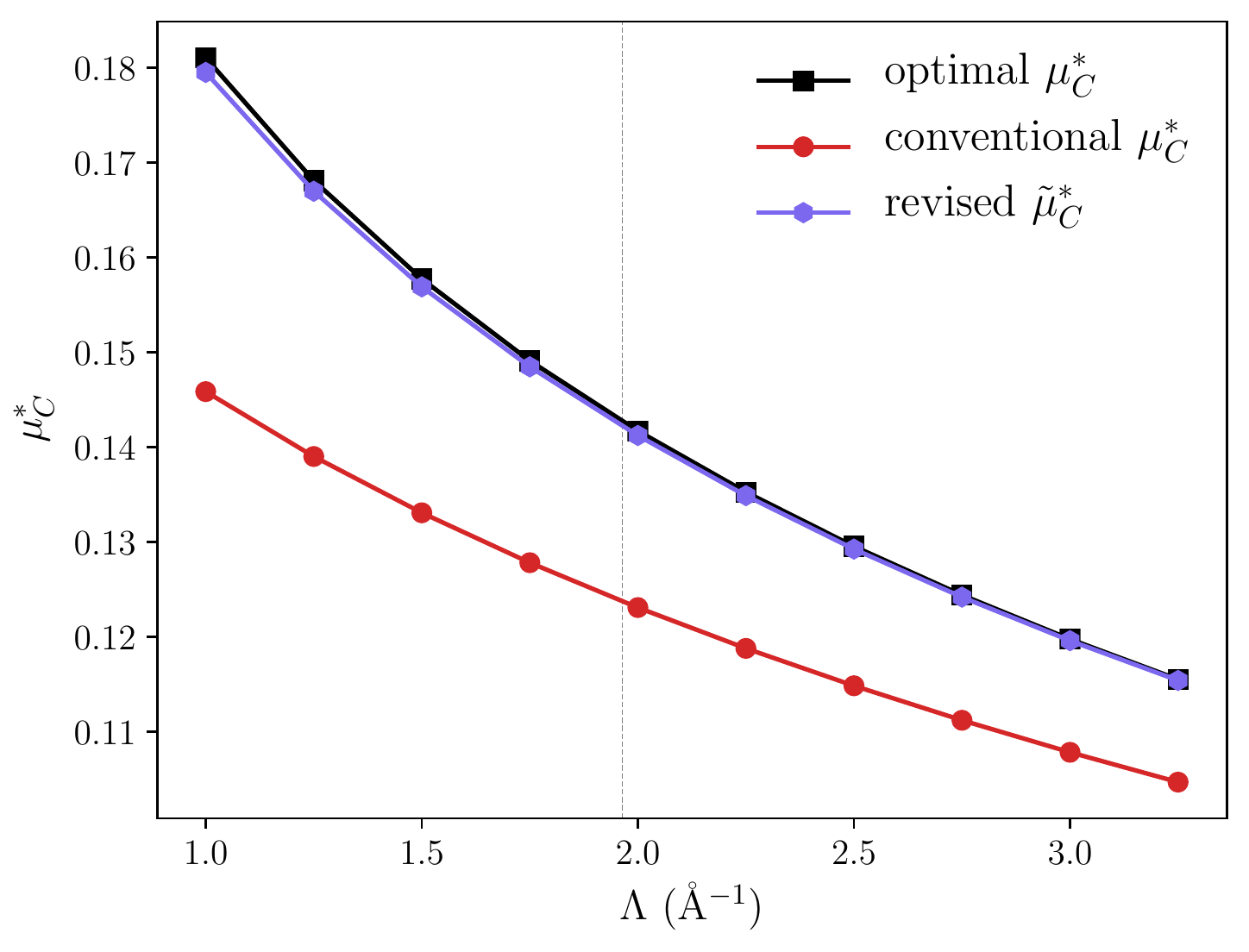}
    \includegraphics[width =\linewidth, keepaspectratio]{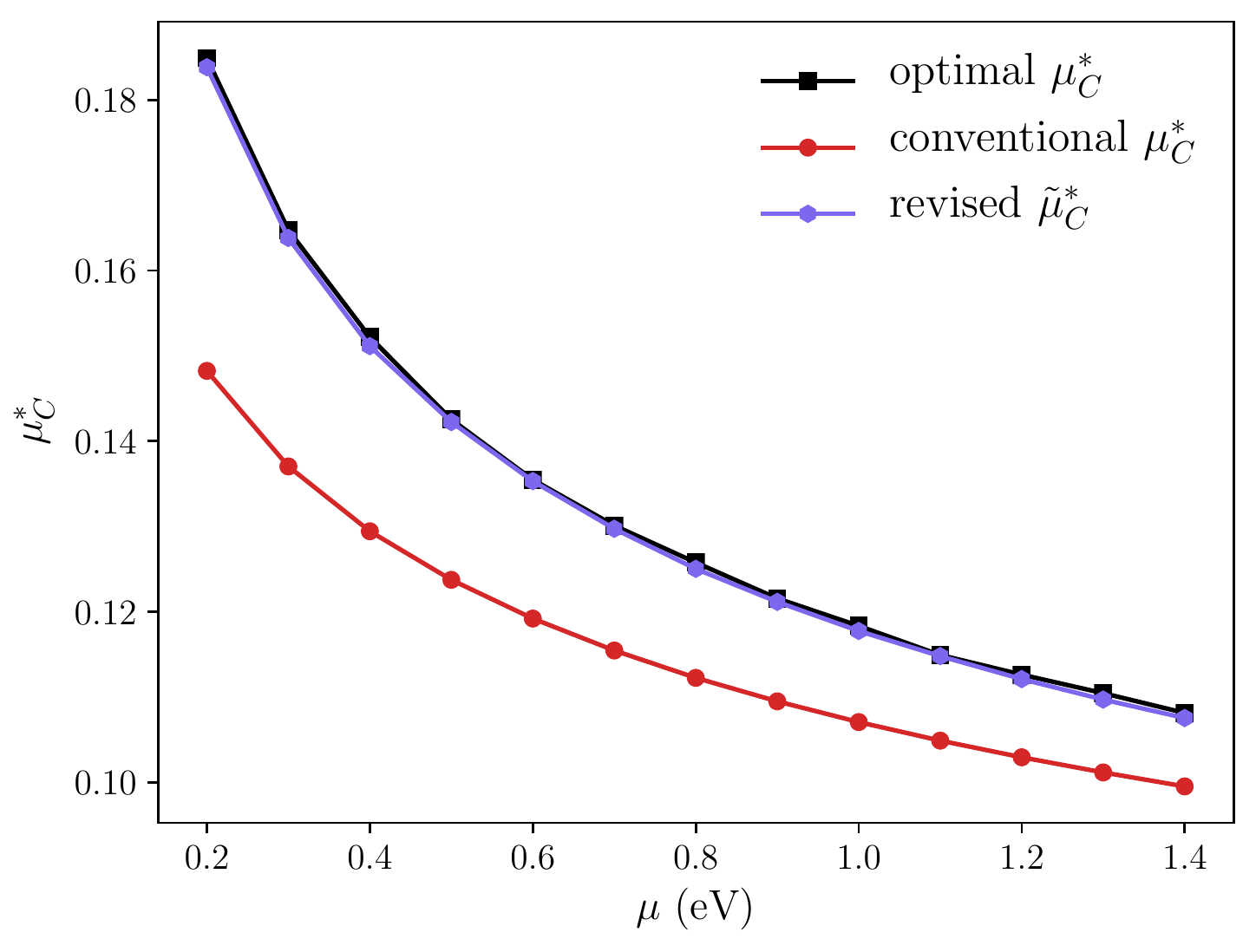}
    \caption{Comparison of the conventional TMA $\mu^*_C$ and the revised $\tilde{\mu}^*_C$ with numerically obtained optimal $\mu^*_C$ as a function of $\Lambda$ (upper panel) and $\mu$ (lower panel). The dashed vertical line in the upper panel denotes $\Lambda_{TF}$.}
    \label{fig:compare}
\end{figure}

In the case of an isotropic dispersion $\xi_q$ and an isotropic Coulomb kernel $V_{kk'}$, we can perform this new approach analytically to obtain an explicit expression for $\tilde{\mu}_C^*$. To this end, we introduce the reduced angle-integrated BSE (at $T=0$), with $z_{kk'} = \frac{f_{kk'}}{U}$ and $z^*_{kk'} = \frac{f^*_{kk'}}{U}$, and project onto the Fermi surface:
\begin{equation}
    z^*_{k_Fk_F}  =  z_{k_Fk_F} - U \int_{\bar{\chi}} dq \, z_{k_Fq}\, \frac{1}{2|\xi_q|} z^*_{qk_F},
\end{equation}
where $f_{kk'}$ and $f^*_{kk'}$ are the angle-integrated versions of $V_{kk'}$ and $W_{kk'}$, respectively (see appendix \ref{app:polar}). To solve this self-consistent equation we choose an ansatz of the form
\begin{equation}
 z^*_{k_Fq} = \alpha z_{k_Fq},
\end{equation}
and obtain
\begin{equation}
    \alpha = \frac{1}{1 + \mu_C \gamma},
    \label{eq:alpha}
\end{equation}
with
\begin{equation}
    \gamma = \frac{1}{\rho_0}\int_{\bar{\chi}} dq \frac{z^2_{k_Fq}}{2|\xi_q|} \, .
    \label{eq:gamma}
\end{equation}
This finally yields the new retarded potential
\begin{equation}
    \tilde{\mu}_C^* = \frac{\mu_C}{1 + \mu_C \gamma}.
    \label{eq:nuStar}
\end{equation}
This result is a generalisation of the TMA potential $\mu^*_C$, which now takes into account the effect of nonlocal screening in the BSE. In the local limit $z_{k_Fq} = 1$ (large $\Lambda$ and large $\mu$) $\gamma$ immediately yields
\begin{equation}
   \gamma =   \int_{\omega_D}^{D} d \xi \frac{1}{|\xi|} = \log\Big( \frac{D}{\omega_D} \Big) \, ,
\end{equation}
such that $\tilde{\mu}_C^*$ correctly reduces to $\mu^*_C$ as in Eq.~(\ref{eq:muStar}). In Fig.~\ref{fig:compare} we analyze the quality of the revised $\tilde{\mu}_C^*$ by comparing it with the optimal $\mu^*_C$, which is obtained by inverting the relation from Eq.~(\ref{eq:deltaL}) using the numerically obtained $\Delta_{k_F}$. Additionally, we show the comparison with the conventional TMA pseudopotential $\mu^*_C$. The qualitative and quantitative agreement between the numerically obtained optimal value and our revised approximation is very good for all $\Lambda$ as well as for different $\mu$, even for small values of $\Lambda$, where the nonlocality of the Coulomb interaction is strong. 

\begin{figure}[htb]
    \centering
    \includegraphics[width =\linewidth, keepaspectratio]{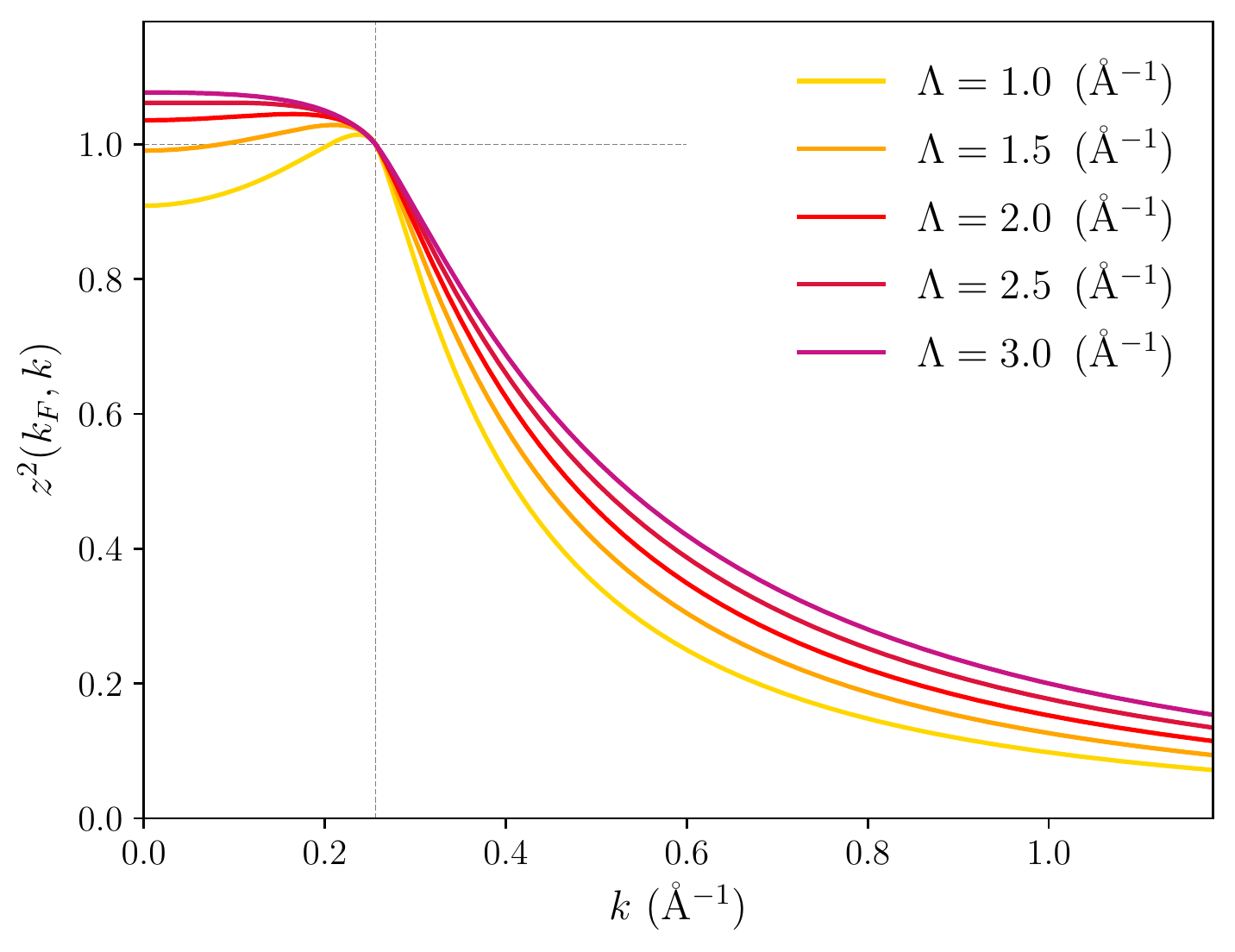}
    \includegraphics[width =\linewidth, keepaspectratio]{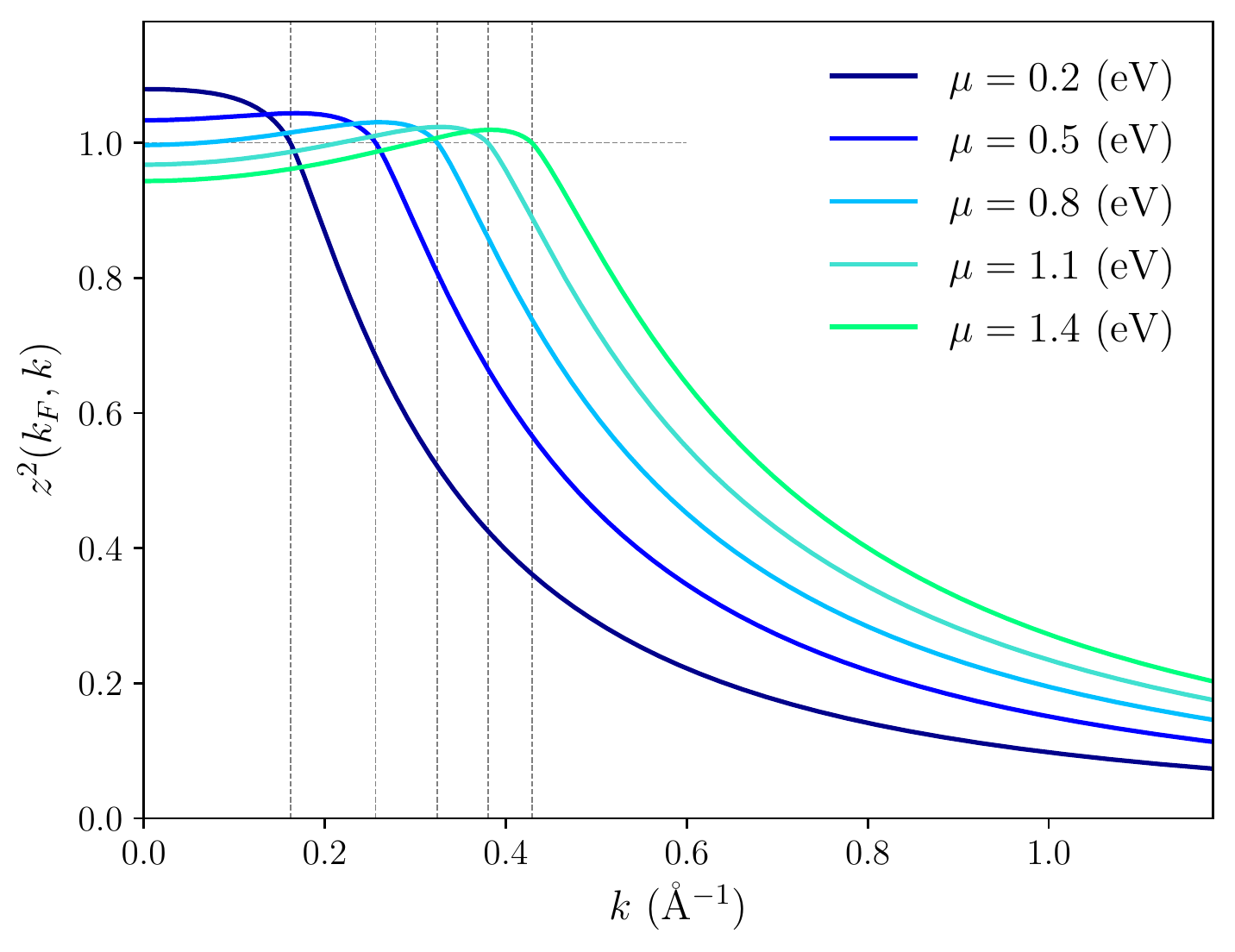}
    \caption{The function $z^2{k_Fk}$ for $\mu =0.5$ eV and different $\Lambda$ (top panel) and for $\Lambda =\Lambda_{TF}$ and different $\mu$ (bottom panel). The vertical dashed lines denote $k_F$.}
    \label{fig:z2}
\end{figure}

In order to understand why $\tilde{\mu}_C^* > \mu^*_C$ we need to analyze $\gamma$ from Eq.~(\ref{eq:gamma}) and therefore the function $z^2_{k_Fk}$. To this end we show in Fig.~\ref{fig:z2} $z^2_{k_Fk}$ for different values of $\Lambda$ and chemical potential $\mu$. We notice that this function is approximately constant for the states below the Fermi level and rapidly decays for the states \emph{above} it. Thus in the presence of nonlocal interactions, pair fluctuations involving occupied sates below the Fermi level contribute to the screening (retardation) as in the case of a local potential, whereas the effects of screening from pair fluctuations involving unoccupied states above the Fermi level are strongly suppressed, as dictated by the decay of $z^2_{k_Fk}$ for large momenta $k$. It is furthermore important to note, that $\gamma$ not only depends on the inverse screening length $\Lambda$, but also strongly depends on the chemical potential $\mu$ through $z^2_{k_Fk}$. This is different to the conventional case, where the chemical potential does not control directly the screening effects. This eventually explains the trends we observed in Fig.~\ref{fig:rel_gap}, where the discrepancy between $\tilde{\mu}_C^*$ and $\mu_C^*$ increases for smaller chemical potential.

\section{Discussion}

Our results show that nonlocal Coulomb interactions can have a strong effect to superconducting properties by, e.g., suppressing $T_c$. In this context the ab initio study on doped layered nitrides by Akashi et al. \cite{akashi_high-temperature_2012} is interesting to note. In this work, the authors derived the superconducting transition temperatures in two ways: (a) via the McMillan-Allen-Dynes (MAD) formula for which they calculate all necessary parameters from ab initio including the \emph{local} retarded TMA Coulomb pseudopotential $\mu^*_C$ and (b) directly via density functional theory for superconductors (SC-DFT) using the full nonlocal Coulomb interaction. Upon neglecting the Coulomb repulsion Akashi et al. find these two approaches to be in good agreement. However, upon taking the Coulomb repulsion into account they consistently find $T_c^{\text{SC-DFT}} < T_c^{\text{MAD}}$. The conventional \emph{local} retarded Coulomb pseudopotential $\mu^*_C$ thus tends to overestimate the critical temperature or, vice versa, the nonlocal Coulomb interaction seems to suppress superconductivity. This observation is fully inline with our findings from above. We expect that using our revised $\tilde{\mu}_C^*$ within the MAD formula will result in $T_c^{\text{SC-DFT}} \approx T_c^{\text{MAD}}$.

In this context experimental data on superconducting transition metal dichalcogenides (TMDCs) is additionally interesting not note. For MoS$_2$ \cite{costanzo_gate-induced_2016, qiu_recent_2021} and NbSe$_2$ \cite{xi_strongly_2015, de_la_barrera_tuning_2018, khestanova_unusual_2018} there is a consistent drop in $T_c$ in their monolayer limits compared to their multilayer compounds. While hybridization and substrate effects might play a role to describe this behaviour \cite{schonhoff_interplay_2016, kamlapure_tuning_2022}, it might (at least partially) also result from the enhanced long-range Coulomb interaction in the monolayer limit, which is suppressed in the multilayer compounds.

\section{Conclusion and outlook}

We analyzed the behaviour of conventional two-dimensional superconductors subject to static nonlocal Coulomb interactions within BCS theory. We found that the nonlocal Coulomb interaction leads to modifications to the superconducting gap function in momentum space, most importantly, in form of a reduced negative amplitude at large momenta away from the Fermi surface. Upon analysing the Bethe-Salpeter equation (BSE) describing the screening of the bare Coulomb repulsion by pair fluctuations, we understood that high-energy screening processes are suppressed in the case of nonlocal Coulomb interactions, which effectively enhances the Coulomb repulsion and thus suppresses superconductivity. 

We demonstrate that the widely applied Fermi-surface averaged Tolmachev-Morel-Anderson \emph{local} Coulomb pseudopotential $\mu^*_C$ overestimates the BSE screening, yielding to too small Coulomb repulsion and thus too large superconducting gaps and transition temperatures as soon as nonlocal Coulomb interactions are present. 
This finding is in line with numerical data by Akashi et al. showing that the TMA $\mu^*_C$ overestimates $T_c$ in layered nitrides \cite{akashi_high-temperature_2012}.

Finally, we re-analyzed the BSE in the presences of nonlocal Coulomb repulsion and were able to derive a revised Tolmachev-Morel-Anderson \emph{local} Coulomb pseudo potential $\tilde{\mu}_C^*$, which takes into account the reduced screening at high energies. This $\tilde{\mu}_C^*$ allows to quantitatively approximate the gap function from the full nonlocal Coulomb kernel at the Fermi level. 
The existence of such a refined \emph{local} pseudopotential is an important finding on its own, as it allows it to be an effective fitting parameter to reproduce experimental data even without microscopic knowledge of the Coulomb interactions. Furthermore, in case one has access to the microscopic Coulomb interaction $V_{kk'}$, our refined $\tilde{\mu}_C^*$ is relatively easy to evaluate.

As in layered materials environmental screening is in general reduced, we expect our findings to be most important for conventional superconductivity in thin film metals in two ways: (1) superconducting properties including transitions temperatures are likely reduced in the thin-film limit due to strong internal nonlocal Coulomb interactions and weak external screening. (2) Environmental or substrate screening to thin film superconductors could be beneficial as it generally reduces the bare Coulomb repulsion in the thin-film superconductor and reduces its long-range character. Thus, environmental screening will effectively decrease the nonlocality of the Coulomb repulsion, which, according to our findings presented here, should enhance $T_c$.

Furthermore, our findings show that nonlocal Coulomb interaction effects are important at reduced Fermi levels. We thus expect the nonlocal Coulomb interaction to reduce superconducting properties mostly in slightly doped layered semiconducting systems with small effective masses. This might be one of the reasons why the critical temperatures in monolayers of doped MoS$_2$ and WS$_2$ are reduced compared to their multilayer counterparts.

\section{Acknowledgement}

We acknowledge fruitful discussions with Tom Westerhout, Yann in't Veld, Andrew J. Millis, Boris V. Svistunov, Nikolay V. Prokof'ev, and Andrey Chubukov. The work of M. I. K. was supported by the European Research Council (ERC) under the European Union’s Horizon 2020 research and innovation programme, grant agreement 854843-FASTCORR.

\appendix

\section{Polar coordinates representation}
\label{app:polar}
    The Coulomb interaction in polar coordinates reads
    \begin{equation*}
    V(k,k', \theta, \theta') = \frac{2\pi e^2}{\Omega}\frac{1}{ \epsilon \sqrt{k^2 + k'^2 -2 kk' \cos(\theta -\theta')} +\Lambda} \, ,
    \end{equation*}
    which allows us to define the angle-integrated interaction as
    \begin{align*}
        f_{kk'} :& =  \int  \frac{d\theta}{2\pi}  \int  \frac{ d \theta}{2\pi}  V(k,k', \theta, \theta') \\
         & = \int  \frac{d\phi}{2\pi} \frac{2\pi e^2/\Omega  }{ \epsilon \sqrt{k^2 + k'^2 -2 kk' \cos \phi} +\Lambda} \, .
    \end{align*}
    The gap equation at $T=0$ then takes the form
    \begin{equation*}
        \Delta_k = \frac{\Omega}{{(2\pi)}} \int dk' k' \big[g_{kk'} - f_{kk'}\big]\, \frac{\Delta_{k'}}{2E_{k'}}.
    \end{equation*}
    In polar representation the conventional Coulomb potential from Eq.~(\ref{eq:muC_2FS}) is given by
    \begin{equation*}
    \mu_C = \rho_0  f_{k_Fk_F} = : \rho_0 U.
    \label{mufinal}
    \end{equation*}

\vspace{5mm}
\section{Analytical Solution of gap equation}
\label{app:sol}
    In the case of a local Coulomb interaction the gap equation from Eq.~(\ref{eq:deltaL}) leads to the system of equations
    \begin{align*}
        \Delta_1 = & (g - U) \int_{\chi} dk \frac{\Delta_1}{2\sqrt{(\xi_k^2 + \Delta_1^2)}}  - U \int_{\bar{\chi}} dk \frac{\Delta_2}{2|\xi_k|} \\
        \Delta_2 = &  U \int_{\chi} dk \frac{\Delta_1}{2\sqrt{(\xi_k^2 + \Delta_1^2)}} - U \int_{\bar{\chi}} dk \frac{\Delta_2}{2|\xi_k|},
    \end{align*}
    which yields after integration
    \begin{align*}
        \Delta_1 = & (g - U) \Delta_1 \rho_0  \mathrm{arsinh} \Big( \frac{\omega_D}{\Delta_1} \Big)  - U \Delta_2 \rho_0  \log \Big(\frac{D}{\omega_D} \Big) \\
        \Delta_2 = &  U  \Delta_1 \rho_0  \mathrm{arsinh} \Big( \frac{\omega_D}{\Delta_1} \Big) - U \Delta_2 \rho_0  \log \Big(\frac{D}{\omega_D} \Big)\, .
    \end{align*}
    This system of equations is solved by the expression from Eq.~(\ref{eq:deltaLSol}).
    
    In the case of nonlocal Coulomb interactions, we know the gap function exhibits a structure in momentum space. However, the domain $\chi$ where $|\xi_k|< \omega_D$ is typically much smaller than $\bar{\chi}$. This allows us to neglect any $k$ dependence of the gap function within the region $\chi$, and approximate it with $\tilde{\Delta}_{k_F} =: \tilde{\Delta}_1$. By denoting the negative part of the energy gap with $\Delta_2(k)$ the angle-integrated gap equation reads:
    \begin{widetext}
    \begin{align}
    \tilde{\Delta}_1 = & g \int_{\chi}dk'\frac{\tilde{\Delta}_1}{2\sqrt{(\xi_{k'}^2 + \tilde{\Delta}_1^2)}}
       - \int_{\chi} dk' f_{k_Fk'}\frac{\tilde{\Delta}_1}{2\sqrt{(\xi_{k'}^2 + \tilde{\Delta}_1^2)}} 
     -  \int_{\bar{\chi}} dk' f_{k_Fk'} \frac{\Delta_2(k')}{2|\xi_{k'}|} \\
    \Delta_2(k) = &  -  \int_{\chi} dk' f_{kk'} \frac{\tilde{\Delta}_1}{2\sqrt{(\xi_{k'}^2 + \tilde{\Delta}_1^2)}} - \int_{\bar{\chi}} dk' f_{kk'} \frac{\Delta_2(k')}{2|\xi_{k'}|}\, .
    \end{align}
    \label{eq:system}
    \end{widetext}
    To proceed we define the dimension-less gap $\delta(k):= \Delta_2(k)/ \tilde{\Delta}_1 $ and search for a solution with the ansatz
    \begin{equation*}
        \delta(k) = -\alpha f_{k_Fk} 
    \end{equation*}
    where $\alpha$ is a scalar. By approximating  $f_{kk'}$ with $f_{kk'_F}$ for $k' \in \chi $ in the first term of the right hand side and employing the isotropy $f_{k_Fk} = f_{kk_F}$, we finally arrive at the expression  
    \begin{equation}
    \alpha = \frac{ \rho_0 \mathrm{arsinh} \Big( \frac{\omega_D}{\tilde{\Delta}_1} \Big)  }{1 +  \frac{1}{f_{k_Fk}}  \int_{\bar{\chi}} f_{kk'} f_{k_Fk'}\frac{1}{2|\xi|} } 
    \label{eq:alpha_gnrl} \, .
    \end{equation} 
    $\alpha$ is here still a function of the external variable $k$, thus a scalar $\alpha$ cannot provide a full solution for all $k$. However, we find: (1) for large values of $k=\bar{k}$, the function $f_{\bar{k}k'}$ becomes approximately constant. This means that we can approximate $f_{\bar{k}k'}$ with $ f_{\bar{k}k_F}$ and obtain an asymptotically correct approximation for $\Delta_2(k)$ in the nonlocal interaction model:
    \begin{equation*}
     \delta({\bar{k}})  = - \frac{\mu_C}{1 + \mu_C \gamma_1} z_{k_F\bar{k}} \mathrm{arsinh} \Big( \frac{\omega_D}{\tilde{\Delta}_1} \Big) \, ,
    \end{equation*}
    where $z_{k_Fk} = \frac{f_{k_Fk}}{U}$ and the coefficient $\gamma_1$ is defined
    \begin{equation*}
    \gamma_1 = \frac{1}{\rho_0}  \int_{\bar{\chi}} dk  \frac{z(k_F, k)}{2|\xi_k|} \, .
    \end{equation*}
     (2) Most importantly, Eq.~(\ref{eq:alpha_gnrl}) allows us to evaluate $\tilde{\Delta}_1$ by estimating the screening (retardation) effects of the high-energy pair fluctuations \emph{acting at the Fermi level}. This is achieved by setting $k=k_F$ in Eq.~(\ref{eq:alpha_gnrl}) such that that we get
    \begin{equation*}
    \alpha =  \frac{ 1 }{1 +   \mu_C \gamma } \rho_0 \mathrm{arsinh} \Big( \frac{\omega_D}{\tilde{\Delta}_1} \Big)
    \end{equation*}
    with
    \begin{equation*}
    \gamma = \frac{1}{\rho_0}  \int_{\bar{\chi}} dk  z^2(k_F, k)\frac{1}{2|\xi_k|}  \, .
    \end{equation*}
    This yields with Eq.(B1)
    \begin{align*}
    1 & =  \Big[  \lambda -   \frac{ \mu_C  }{ 1 +   \mu_C \gamma } 		\Big] \mathrm{arsinh}\Big( \frac{\omega_D}{\tilde{\Delta}_1} \Big) \, ,
    \end{align*} 
    which allows us to identify the revised TMA pseudopotential $\tilde{\mu}_C^*$ as
    \begin{equation*}
    \tilde{\mu}^*_C =  \frac{ \mu_C  }{ 1 +   \mu_C \gamma }  \, .
    \end{equation*}
    This result is consistent with the one we obtained by analyzing the Bethe-Salpeter equation.

\end{document}